\documentclass[pdftex,a4paper,preprint]{revtex4}

\usepackage[english]{babel}
\usepackage{times}

\usepackage[unicode=true, pdfstartview=FitH]{hyperref}

\usepackage[]{graphicx}

\newcommand{\Ethr}{E_{\text{thr.}}}

\begin{document}

\title{Spati-temporal distribution of cascade particles below the maximum of EAS development with $E_{0} \ge 10^{17}$~eV}

\author{Stanislav~Knurenko}\email[1]{s.p.knurenko@ikfia.ysn.ru}
\author{Zim~Petrov}\email[2]{pze@ikfia.ysn.ru}
\author{Yuri~Yegorov}\email[3]{yuriyegorov@ikfia.ysn.ru}
\author{Nikolai~Dyachkovsky}\email[4]{nikolayakut@rambler.ru}

\affiliation{Yu.~G.~Shafer Institute for cosmophysical research and aeronomy SB RAS}

\begin{abstract}
Signal from scintillation detectors in showers with energies $10^{17} - 10^{20}$~eV has been analyzed. It was determined, that for core distances $\ge 200$~m the time scale of a particle arrival time at the detector constitutes $40 - 5000$~ns and longer. Events with fine structure of the signal have been observed. The pulse consists of several peaks (for vertical showers) and a single pulse (for inclined showers). There are some anomalous events with two pulses wherein such a pulse shape is presented in two, three and more detectors. A period between these pulses makes $60 - 350$~ns.
\end{abstract}

\maketitle

\section{Introduction}

For many years at the Yakutsk EAS array studies of the pulse shape from Cherenkov and scintillation detectors have been performed~\cite{ref1, ref2}. Initial measurements of the pulse shape from $2$-m$^{2}$ scintillation detector at the Yakutsk array have been performed in 1975, continued in 1988-1990 and renewed in 2005. Lately, a special designed setup has been used for this purpose, which consists of scintillation and Cherenkov detectors and quick-response electronics. All this allowed reconstruction of time-related characteristics of forward and back fronts and curvature and thickness of shower disk with a good precision. Measurements have shown that time-base of the signal in scintillation detector can be used in quantative analysis of EAS data, including cosmic ray mass composition data.

\section{Equipment used for registering the pulse shape from charged particles}

For measurements at the Yakutsk EAS array scintillation detectors of different areas and thickness were used. Two detectors were 2~m$^{2}$ and were separated by $51$~m from each other, the rest six were $1$, $0.25$, $0.10$ and $0.02$~m$^{2}$ correspondingly and were placed in apexes of a tetragon with sides $7$ and $4.5$~m. One of the detectors ($s = 0.10$~m$^{2}$) had a lead ceiling (thickness $10$~cm), similar detector had no top and was covered with thin black paper. These detectors were appointed for studying the influence of the cover material on detector response and the ratio between muons with $\Ethr \ge 0.3$~GeV and electrons in a shower. Scintillation detector with area of $1$~m$^{2}$ was $1$~cm thick and effectively registered particles with $\Ethr \ge 2$~MeV. For studying the spati-temporal characteristics of Cherenkov radiation a camera obscura was used~\cite{ref3}.

\begin{figure}
\centering
\includegraphics[width=0.85\textwidth]{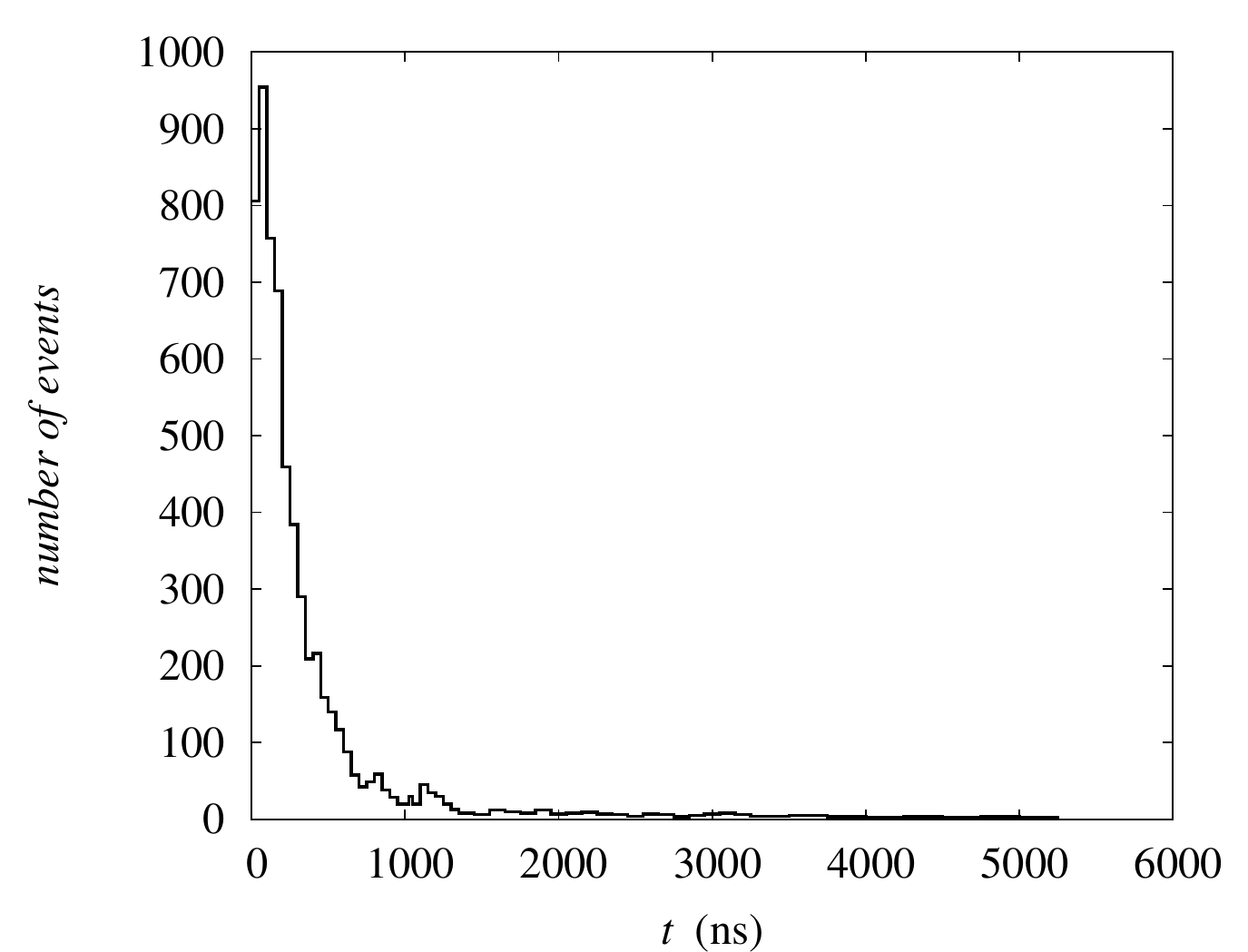}
\caption{Delayed pulses distribution (relative to the ``fastest'' particle) over delay time $t$. Considered core distances in the range of $80 - 1500$~m}
\label{fig1}
\end{figure}

\subsection{Recording system and control}

For pulse shape recording we used an industrial computer of increased reliability equipped with integrating board with 19 PCI slots. {\em La-n10-m8} boards with two fast 8-bit AD-converters with $100$~MHz sampling rate and $2$~Mb buffer store were plugged in these PCI slots. The computer is meant for 20 ADCs. Registration is controlled by external ``masters''. Masters generated by the main array when signals from three noncollinear scintillation detectors spaced by~500~m coincide (so-called {\em Trigger-500}). Masters are generated by the small Cherenkov array when signals from three integral Cherenkov detectors located in apexes of equilateral triangle with side $50$, $100$ and $250$~m coincide. After the registration program start, ADCs continuously convert signals from detectors outputs and repeatedly record them into the area of buffer memory called ``pre-history''. In the ``pre-history'' the latest data on digitization of valid signals are stored, period between neighbouring counts is $10$~ns. After accepting the ``master'', signals of ``master'' coloring and data on the amplitude of calibration LED are recorded into the area of buffer memory called ``history''. Accepted frames of events are formed as a file record and added to the initial file, created automatically at the beginning of observation. Such a selection system allows pulse registration in showers with energies $10^{15} - 10^{19}$~eV.

\begin{figure}
\centering
\includegraphics[width=0.85\textwidth]{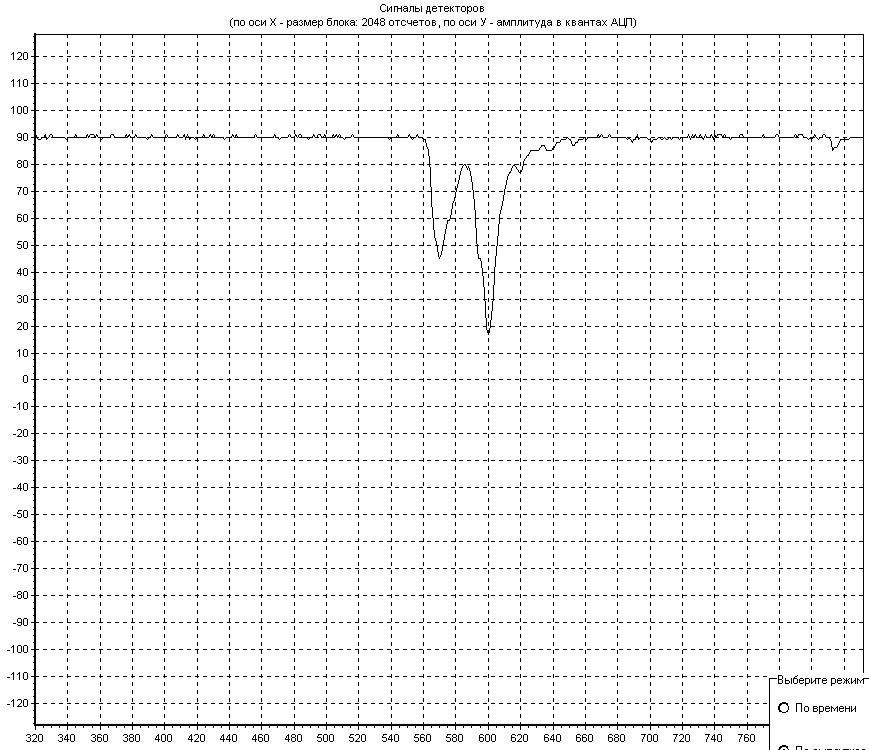} 
\caption{An example of anomalous pulse shape. Channel number~16 (detector without ceiling). There are also double pulses in other detectors but with very small amplitude}
\label{fig2}
\end{figure}

\section{Registration results}

After $\sim 18000$~hours of array duty cycle $\sim 1.2 \times 10^{5}$ showers were registered with energy above $10^{15}$~eV. In every event arrival time of forward front of charged particles and Cherenkov light, build-up time and pulse half-width were measured and pulses from delayed particles were recorded on a time scale up to $15$~mcs. Shower characteristics such as primary energy, zenith and azimuth angles, shower size (full number of charged particles, full number of muons with $\Ethr \ge 1$~GeV) and depth of maximum development were estimated with scintillation and Cherenkov detectors of the main array.

During the analysis several peculiarities were discovered in the shape of registered pulses as well as in their time distribution. This concerns both ``young'' (vertical) showers and ``old'' (inclined) showers. The former have half-width of the pulse $200-250$~ns, the latter~---$150-180$~ns. There are showers with large amount of delayed particles and in some events  the time scale notably extends up to $10$~mcs.

As it follows from fig.~\ref{fig1}, in most EAS events delayed particles are distributed over the interval $\Delta t = 40 - 2000$~mcs, which coincides with particles picking time of ADC at the Yakutsk array.

\begin{figure}
\centering
\includegraphics[width=0.85\textwidth]{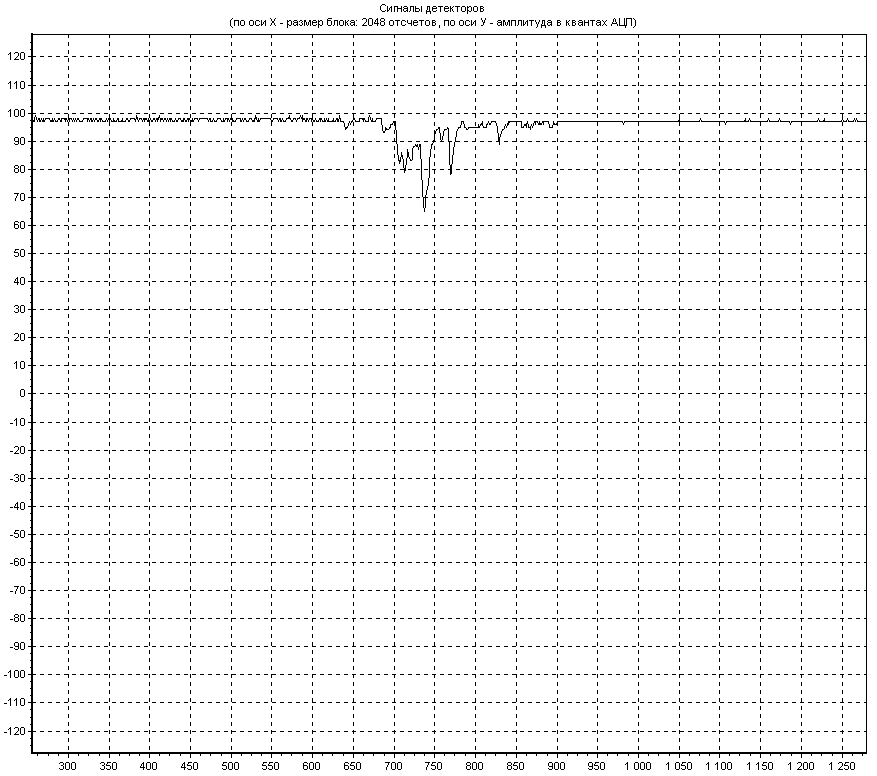}
\caption{Pulse shape recorded at the core distance $R = 1298$~m. $E_{0} = 1.7 \times 10^{19}$~eV, zenith and azimuth angles equal to $18^{\circ}$ and $78^{\circ}$ respectively}
\label{fig3}
\end{figure}

From the huge amount of data one can distinguish a small group of showers with clearly visible double-peaked shape (see fig.~\ref{fig2}). Such a shape is traced at a single detector, as well as at two-three detectors. The delay time of the second group of particles lies within $60 - 350$~ns from the first one. The number of such events is all in all $\sim 1.5 - 2$\% from the total number of registered EAS events. From all the data listed above, it is possible to classify showers by their pulse shape. Electron-photon shower has larger pulse half-width and half-height compared to ``muonic'' one. Pulse structure consists of many saw-like peaks distributed in interval from $0$ to $1000$~ns starting with the first particle arrived at the detector (see fig.~\ref{fig3}). Such pulses to a greater degree reflect string nature of shower particles generating including electron-photon component of the shower. Such a pulse shape usually is peculiar to vertical showers with low maximum in the atmosphere.

\begin{figure}
\centering
\includegraphics[width=0.85\textwidth]{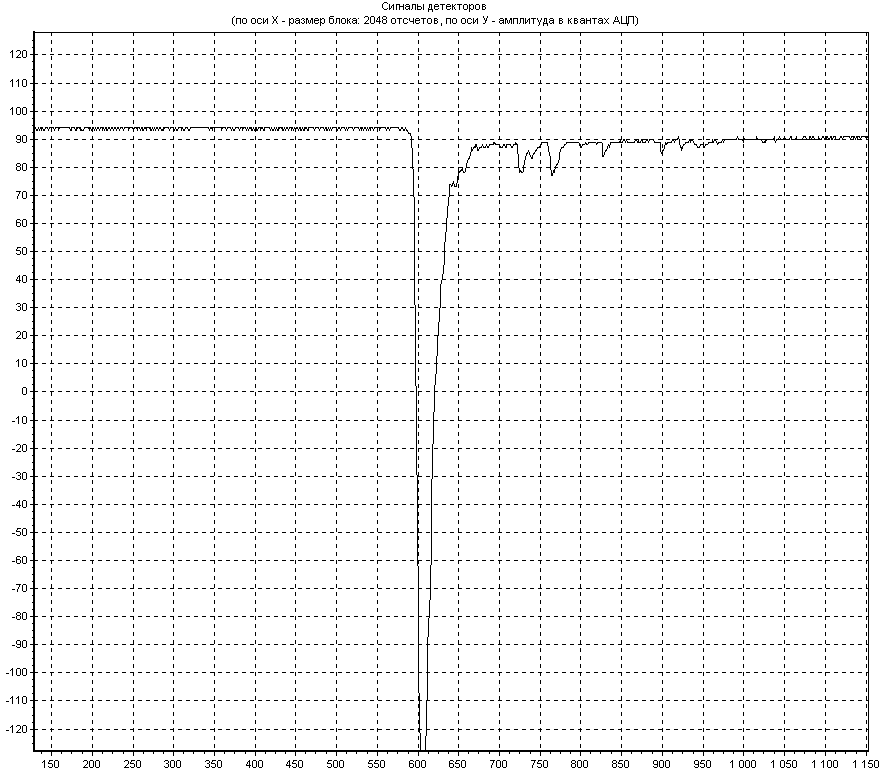}
\caption{Pulse shape recorded at the core distance $R = 1000$~m. $E_{0} = 2.4 \times 10^{19}$~eV, zenith and azimuth angles equal to $56^{\circ}$ and $200^{\circ}$ respectively}
\label{fig4}
\end{figure}

``Muonic'' showers have round pulse peak and notably $30$\% lesser half-width compared to pulse from the electron-photon shower component (see fig.~\ref{fig4}). Judging by the pulse such particles arrive compactly i.e. distributed in lesser time interval from $0$ to $550$~ns. Such a pulse shape can be observed in inclined showers and in showers with low maximum of development.

It has been assumed since the very beginning that both single detectors and their combinations are to be considered in the analysis. This is connected with the fact, that detectors have different relative apertures on core distances and different threshold for registering relativistic particles~--- $10$, $5$ and $2$~MeV. Also, the relation of signals from two geometrically identical detectors, with and without additional shielding ($\Ethr \ge 0.3$~GeV for muons) was taken into account. Preliminary analysis of the experimental data has shown that scintillation detectors with different areas have different registering efficiency. In showers with $E_{0} = 10^{17} - 10^{18}$~eV counters with $s = 0.1, 0.25$~m$^{2}$ effectively operate at core distance $R \le 500$~m and counter with $s = 2$~m$^{2}$ does so at $R \le 1500$~m. Thus, smaller counters can participate in the ``trigger-500'' generating and the larger ones~--- in the ``trigger-1000'' generating. It has been noted, that pulse characteristics, build-up time of the forward front, half-width on the half-height and response strongly depend on the detector type (geometry, method of light gathering~--- direct or reflected) and on the type of voltage divider on PMT's dynodes. It was shown by comparing the responses from fundamentally different detectors with $s = 1$~m$^{2}$ and $s = 2$~m$^{2}$. Reaction of the scintillation counter with $s = 1$~m$^{2}$, gathering light with optical fiber mounted into the plastic scintillator and with fast PMT ``FEU-115'' is $200 - 300$~ns faster than that of the large scintillation counter with $s = 2$~m$^{2}$ and this fact makes it preferable for precise measurements of time-related characteristics of a shower. From the analysis of showers of different energies one may conclude, that the signal in showers with energy $\sim 10^{17}$~eV is represented by the single-peaked pulse, not unlike the pulse from inclined shower and this fact speaks well for faster development compared to showers with energy $\sim 10^{19}$~eV with maximum of development at the depth $750 - 850$~g/cm$^{2}$.

\begin{figure}
\centering
\includegraphics[width=0.85\textwidth]{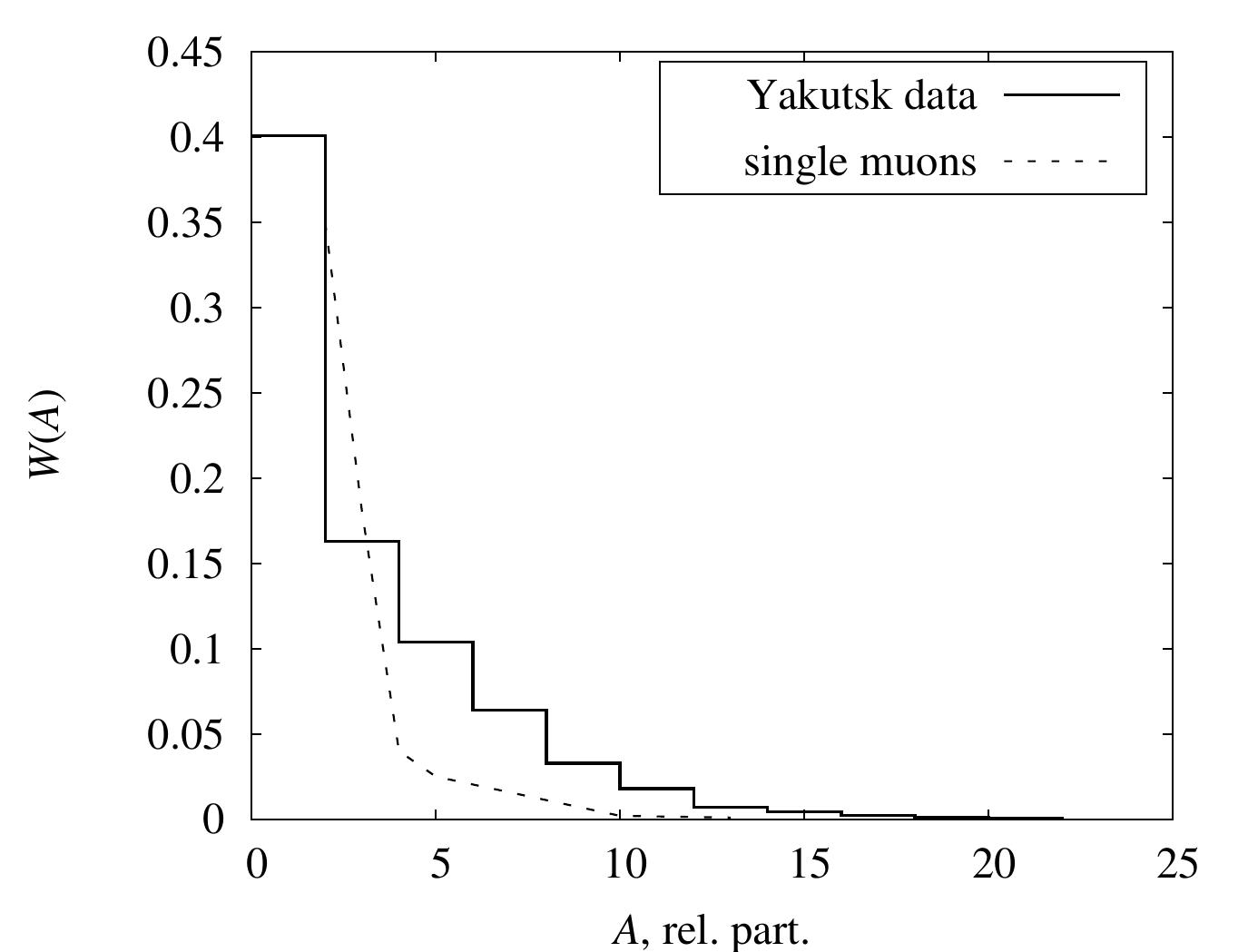}
\caption{Delayed pulses distribution over the amplitude $A$. Histogram~--- the Yakutsk EAS array data. Curve~-- single muons (QGSjet~\cite{ref5}). (It should be noted, that there are showers, where response from a shielded detector with $\Ethr \ge 0.3$~GeV is bigger than that from an unshielded one)}
\label{fig5}
\end{figure}

Analysis of the signal amplitude also has shown interesting results (see fig.~\ref{fig5}). To begin with, in some cases the amplitude in shielded detectors is much higher, than in unshielded ones. Preliminary analysis of such showers ($E_{0} \sim 10^{17} - 10^{18}$~eV, core distance $R \le 200$~m) points to the presence of low-energy hadrons in a stream, which interact with shielding material and generate a micro-shower resulting in growth of the signal amplitude. Secondly, from the comparison of experimental amplitudes distribution with QGSjet simulation for single muons~\cite{ref5} it follows that in the range of $n \ge 5$ relativistic particles, there is an excess of registered particles over simulated, which also requires explanation. One of the versions~--- generating of Cherenkov radiation in glass of the PMT's photocathode, which we observed in special experiments with closed Cherenkov detectors at $R \le 125$~m from the core in showers with energies $10^{17} - 10^{19}$~eV. It should be stressed out, that the portion of such events is rather small and makes about $3$\%.

\subsection{Forward and rear EAS front. Thickness and curvature of the shower disk.}

\begin{figure}
\centering
\includegraphics[width=0.85\textwidth]{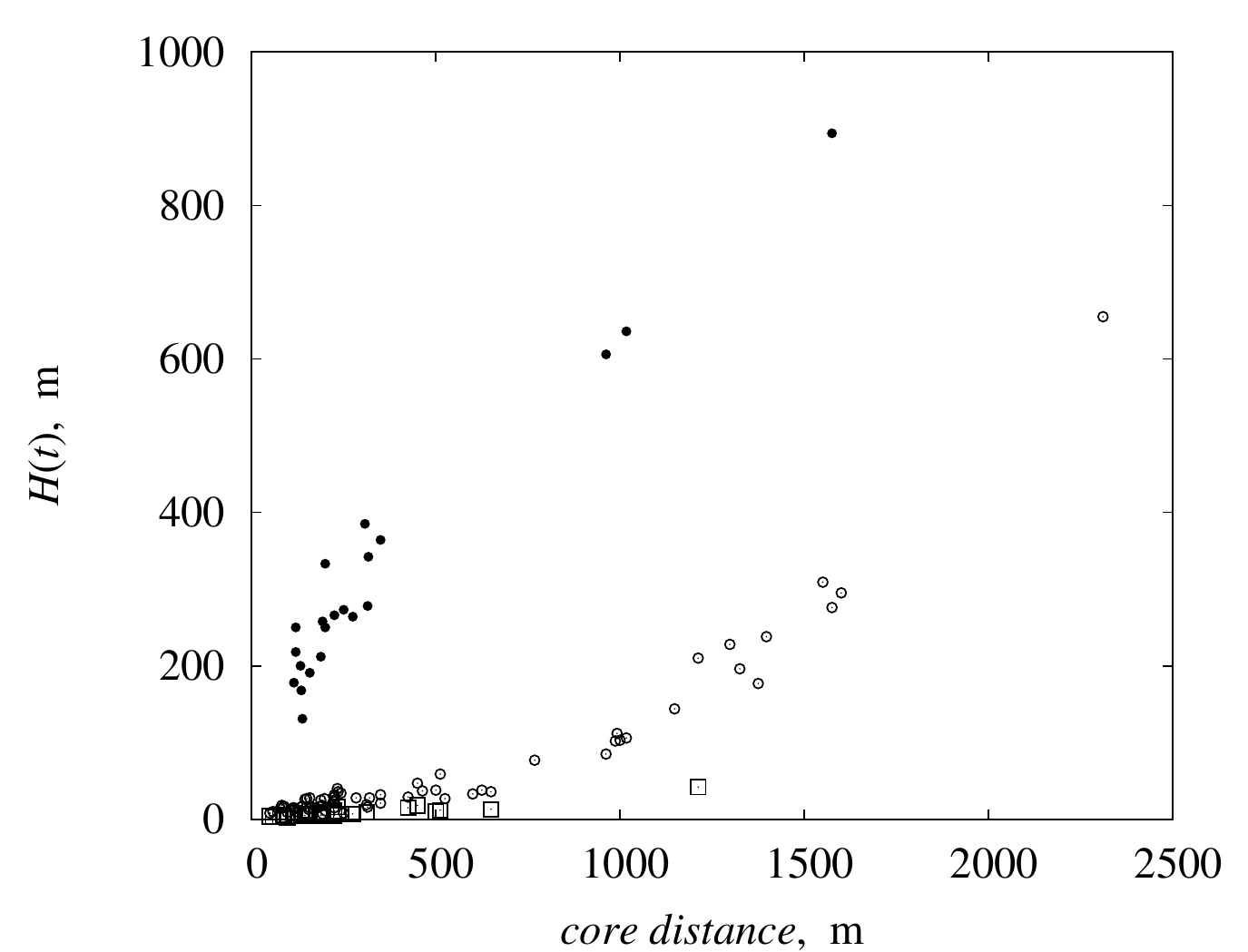}
\caption{Forward and back EAS fronts reconstructed by time delays measured in showers with energies above $10^{16}$~eV. Open circles~--- forward front (charged particles), filled circles~--- rear front. Squares~--- Cherenkov photons}
\label{fig6}
\end{figure}

On fig.~\ref{fig6} time-related shower characteristics are shown, obtained in measurement of charged particles and Cherenkov photons arrival times at the array detectors. A wide distance range was covered, showers with energy above $10^{17}$~eV were selected. Measurements have shown, that up to distances $\sim 300$~m forward front is semi-flat and slightly exceeds the precision of zenith angle measurement at the Yakutsk array ($100$~ns). At $R \ge 300$~m from the core front delay increases significantly and at $1000 - 1500$~m amounts $400 - 800$~ns, requiring taking it into account in measurement of EAS arrival angles.

Using the value of particle arrival delay at the detector $\left< \tau \right>$, one can obtain curvature radius of the shower front. These times characterize areas of shower development in the atmosphere from which at a given core distance the main portion of particles arrives. Radius of curvature is determined by formula:
\begin{displaymath}
R_{\text{curv}} = \frac{R^{2} - (c \tau)^{2}}{2 c \tau}\text{,}
\end{displaymath}
where $R$~--- core distance, $\tau$~--- particle delay, $c$~--- speed of light. For experimental data of the Yakutsk array at mean core distance $790$~m and $\left< \tau \right> = 248$~ns, $R_{\text{curv}} = 4180$~m. Along with the core distance the curvature radius increases, meaning that particles arrive from larger heights.

\section{Conclusion}

Registration of time-related shower characteristics (particle arrival time at the detector, pulse shape in scintillation detectors) allowed us to make the following conclusions:

\begin{enumerate}

\item
``Young'' and ``old'' showers have drastically different pulse shape.

\item
Pulses delayed by $\ge 2$~mcs are observed in showers but not in all. Anomalous pulse shapes are presented.

\item
Curvature of the forward front of showers with energy $> 10^{17}$~eV at the core distance $500 - 800$~m equals $3.9 - 4.9$~km.

\item
Analysis of the response from scintillation detectors with different energy thresholds have shown that pulses delayed by larger periods are generated by low-energy particles (electrons) which were possibly born in the process of neutron moderation in the frozen ground~\cite{ref6}.

\end{enumerate}


\begin{thebibliography}{99}

\bibitem{ref1}
M.~N.~Dyakonov et al. ``Ultra-high energy cosmic rays''. Yakutsk, YaF SO AN SSSR. 15-33, 1979.

\bibitem{ref2}
V.~P.~Artamonov, S.~P.~Knurenko, V.~A.~Kolosov et al. Registration of the pulse shape from charged particles at the Yakutsk EAS array. CRC Poc. Suppl. (part~1). Alma-Ata. 33-34, 1989.

\bibitem{ref3}
Z.~Ye.~Petrov, S.~P.~Knurenko, I.~Ye.~Sleptsov et al. ``Sovremennye problemy kosmicheskoi fiziki'' Proc. Suppl. Yakutsk, 2008, p.87--90.

\bibitem{ref5}
V.~B.~Atrashkevich, O.~V.~Vedeneev, K.~K.~Garipov et al. Izv.~AN, ser. fiz., v.~58, No.~12, 1994, p.98--102.

\bibitem{ref6}
A.~D.~Erlykin, EAS and the physics of high energy interactions. The talk at 20-th ECRS, Lissboa, 2006.

\end{thebibliography}
\end{document}